\documentstyle[multicol,epsfig,aps]{revtex}

\begin{document}

\title{Deposition of magnetic particles: A computer simulation study}

\author{F. de los Santos$^{1,2}$, M. Tasinkevych$^2$, J. M. Tavares$^{2,3}$
and P. I. C. Teixeira$^{4,5}$}

\address{$^1$Physics Department, Boston University \\
590 Commonwealth Avenue, Boston, MA 02215, USA \\
$^2$Centro de F\'{\i}sica Te\'{o}rica e Computacional
da Universidade de Lisboa \\
Avenida Professor Gama Pinto 2, P-1649-003 Lisbon, Portugal \\
$^3$Departamento de Ci\^{e}ncias Exactas e Tecnol\'{o}gicas, 
Universidade Aberta \\
Rua Fern\~{a}o Lopes 9, $2^{\underline\circ}$ D$^{\underline{\rm to}}$, 
P-1000-132 Lisbon, Portugal\\
$^4$Faculdade de Engenharia, Universidade Cat\'{o}lica Portuguesa \\
Estrada de Tala\'{\i}de, P-2635-631 Rio de Mouro, Portugal \\
$^5$Departamento de Engenharia de Materiais \\
and Instituto de Ci\^{e}ncia e Engenharia de Materiais e Superf\'{\i}cies  \\
Instituto Superior T\'{e}cnico \\
Avenida Rovisco Pais, P-1049-001 Lisbon, Portugal}

\date{30 September 2002}

\maketitle

\begin{abstract}
We report a Monte Carlo simulation of deposition of magnetic particles 
on a one-dimensional substrate. Incoming particles interact with those 
that are already part of the deposit via a dipole-dipole potential. 
The strength of the dipolar interaction is controlled by an effective 
temperature $T^*$, the case of pure diffusion-limited deposition 
being recovered in the limit $T^*\to\infty$.
Preliminary results suggest that the fractal dimension of the deposits 
does not change with temperature but that there is a (temperature-dependent) 
cross-over from regimes of temperature-dependent to universal behaviour.
Furthermore, it was found that dipoles tend to align with the local 
direction of growth.

\vspace{0.4cm}
\end{abstract}

\begin{multicols}{2}

\section{Introduction}

\label{sec-intro}

Magnetic particles are a key ingredient of many modern data 
recording and storage devices, from music tapes to computer hard 
disks \cite{Hadjipanayis:2001}. For these applications smooth, regular 
magnetic layers are usually desired. However, dipoles, both magnetic 
and electric, display a fondness for arranging themselves into highly 
inhomogeneous structures. This is a consequence of the very strong 
anisotropy of the dipole-dipole interaction, which couples the 
orientations of the dipole moments with that of the interparticle 
vector. Because the potential energy at a fixed separation is lowest 
for a head-to-tail geometry, chain formation is particularly favoured 
in ferrofluids (dispersions of ferromagnetic particles) 
\cite{Rosensweig:1985} or electrorheological fluids (dipersions
of highly polarisable particles in solvents with low dielectric constant)
\cite{Halsey:1992} in magnetic and electric fields, respectively. Whether
such chaining can occur in zero field in the absence of any interactions
other than dipolar, is experimentally uncertain; it has been seen in 
simulations, but is not yet fully accounted for theoretically. 
Likewise, what is the true equilibrium structure of a solid of 
hard magnetic particles is still unsettled 
(see, e.g., \cite{Teixeira:2000} and references therein).

It is therefore of great importance, from a practical as well as from a 
fundamental point of view, to investigate the influence of true long-range
magnetic dipolar forces on the geometry of particle aggregates, so as to be 
able to exert better control over them. This is especially relevant to the 
very novel field of self-assembled nanostructured magnetic materials,
where the aim is to allow different microscopic components to organise 
themselves into complex functional patterns once their interactions have 
been appropriately tailored \cite{Hadjipanayis:2001,Himpsel:1998}. 
Many of these devices, either existing or at the design stage, have 
low dimensionalities (e.g., wires and films), at which simulations 
of model systems have revealed the chaining tendency to be particularly 
strong \cite{Weis:1998,Weis:2002,ZM:2002a,ZM:2002b}.

To our knowledge, there is detailed only one study of how dipolar 
interactions alone affect growth. 
Pastor-Satorras and Rub\'{\i} \cite{Rubi:1995} simulated an 
off-lattice, two-dimensional particle-cluster aggregation model. They 
found a monotonic variation of the fractal dimension of the aggregates 
as a function of temperature (i.e., dipole strength), the limit of 
diffusion-limited aggregation (DLA) being recovered at high temperatures.
In addition, a separate investigation by the same authors has shown that 
highly structured layers could be obtained at low temperatures by 
sequential adsorption of dipolar particles \cite{Rubi:1998}.
See \cite{Newref1,Newref2} for related work on other systems.

Here we report on a simulation of dipole deposition on a one-dimensional
(1d) substrate (i.e., a line). In the limit of zero magnetic moment this 
reduces to diffusion-limited deposition (DLD), which should exhibit the 
same geometrical properties as DLA \cite{Meakinbook}; our work thus serves 
as a both a check and an extension of Pastor-Satorras and Rub\'{\i}'s to 
the case of an infinite (in one spatial dimension) system. In 
particular, we want to ascertain: firstly, whether the fractal 
dimension actually changes owing to the strong anisotropy of the 
dipolar interaction: and, secondly, what is the correlation between 
the orientations of dipoles in the aggregate and its direction(s) of 
growth. For computational convenience our dipoles are restricted to 
reside at the sites of a two-dimensional (2d) square lattice (although 
they can point in any direction of three-dimensional (3d) space): we 
assume that any effects coming from this discretisation of space are much 
smaller than those of the interparticle potential. That this should be
so is suggested by results for DLA \cite{Meakin:1986} (but remains of 
course to be confirmed by full off-lattice calculations). Furthermore, our
analysis in terms of the concepts of fractal geometry presumes that {\it all}
our deposits are self-similar over some lengthscale larger than the mesh 
spacing but smaller than the deposit size \cite{Meakinbook}; again, this 
need not be true of the smallest deposits, but these contain only a very 
small fraction of the total number of particles.

The present paper is a natural progression to non-equilibrium processes 
from our earlier researches on the thermodynamics and phase equilibria of 
dipolar fluids \cite{Teixeira:2000}.
It is organised as follows: in section \ref{sec-meth} we describe
our model and the simulation method employed. Then in section \ref{sec-res}
we present and discuss our results, specifically comparing them with those 
of Pastor-Satorras and Rub\'{\i} \cite{Rubi:1995}. Section \ref{sec-disc} 
summarises our findings and outlines prospects for future research.
Technical details pertaining to the treatment of the long range of
the dipole-dipole interaction are relegated to two appendices.

\section{Model and simulation method}

\label{sec-meth}

Our simulations were performed on a (1+1)-dimensional square lattice
of width $L=800a$ sites and any height that can accommodate $N$ dipoles, 
where $a$ is the mesh spacing and the adsorbing substrate coincides with 
the bottom row (henceforth we take $a=1$). 
Periodic boundary conditions are imposed in the
direction parallel to the substrate. Each particle carries
a 3d dipole moment of strength $\mu$
and interact through the pair potential
\begin{equation}
\label{dipol_inter}
\phi_D(1,2) = -\frac{\mu^2}{r^3_{12}}
\left[ 3({\mbox{\boldmath$\hat\mu_1$}} \cdot {\bf\hat r_{12}}) 
({\mbox{\boldmath$\hat\mu_2$}} \cdot {\bf\hat r_{12}}) 
-{\mbox{\boldmath$\hat\mu_1$}} \cdot {\mbox{\boldmath$\hat\mu_2$}}\right],
\end{equation}
where $r_{12}(\geq a)$ is the distance between particles 1 and 2, 
${\bf\hat r_{12}}$ is the two-dimensional (2d) unit vector along 
the interparticle axis, and ${\mbox{\boldmath$\hat\mu_1$}}$ and 
${\mbox{\boldmath$\hat\mu_2$}}$ are the 3d unit vectors in the 
direction of the dipole moments of particles 1 and 2 respectively. 
Finally, `1' and `2' denote the full set of positional and 
orientational coordinates of particles 1 and 2.

A particle is introduced at a lattice site $(x_{in},H_{max}+A\,L)$,
where $x_{in}$ is a random integer in the interval $[1,L]$, $H_{max}$
is the maximum distance from the substrate to any particle in the 
deposit, and $A$ is a constant. The dipole moment of the released 
particle is oriented at random. The particle then undergoes a random walk 
by a series of jumps to nearest-neighbour lattice sites, while
experiencing dipolar interactions with the particles
that are already attached to the deposit. We incorporate the 
effects of these interactions through a Metropolis algorithm. If the 
deposit contains $N$ particles, then the interaction
energy of the $(N+1)$th incoming particle (the random walker) with the 
particles in the deposit is given by 
$E({\bf r},{\mbox{\boldmath$\hat\mu$}})=\sum_{i=1}^N
\sum_{\mbox{\boldmath$n$}}
\phi_D(i,N+1)$, where ${\bf r}$ and ${\mbox{\boldmath$\hat\mu$}}$ 
are the current position and the 
dipole orientation of the random walker respectively
($ {\bf r} $ is a 2d vector). Then we randomly choose a new position 
${\bf r^{\prime}}$ ($|{\bf r} -{\bf r^{\prime}}| =a$) and a new dipole 
orientation ${\mbox{\boldmath$\hat\mu^{\prime}$}}$ for the random walker; 
this displacement is accepted with probability
\begin{equation}
\label{metropolis}
p={\rm min}\left\{1,\exp\left[- \frac{E({\bf r^{\prime}},
{\mbox{\boldmath$\hat\mu^{\prime}$}})
-E({\bf r},{\mbox{\boldmath$\hat\mu$}})}{T^*} \right]\right\}.
\end{equation}
$T^*=k_BTa^3/\mu^2$ is an effective temperature, 
inversely proportional to the dipolar energy scale. In the limit 
$T^*\to 0$ only displacements that lower the energy $E({\bf 
r^{\prime}},{\mbox{\boldmath$\hat\mu^{\prime}$}})$ are accepted. On the 
other hand, in the limit $T^*\rightarrow \infty$ all displacements are 
accepted and our model reduces to the well-known DLD \cite{Meakinbook}.

The long range of the dipole-dipole
interaction was treated by the Ewald sum method (see Appendix 
\ref{sec-Ewald} for details). In our simulations we set $\alpha=10/L$,
for which it suffices to retain terms with $n=0$ in the real space sum of
equation (\ref{eq:energy3}). The sum in reciprocal space extends over all
lattice points $k=2\pi n/L$ with $|n|\leq 16$, whereas the sum in real
space is truncated at $L/2$.

The particle eventually either contacts
the deposit (i.e., becomes a nearest neighbour of another particle that
is already part of the deposit) or moves away from the substrate.
In the latter case, if the particle reaches a distance from the substrate
greater than $H_{max}+2A\,L$, it is removed and a new one is launched.
Once a particle has reached the substrate or the deposit, its dipole
relaxes along the direction of the local field created by all other
particles in the deposit.
In all simulations reported here we took $A=1$; larger values of $A$ 
were tested and found to give the same results, but with drastically 
increased computation times.

\section{Results and discussion}

\label{sec-res}

Four effective temperatures were considered:
$T^*=10^{-1}$ (28 deposits), $10^{-2}$ (41 deposits), $10^{-3}$ 
(42 deposits) and $10^{-4}$ (54 deposits), chosen to be in the range
where the fractal dimension of dipolar DLA clusters is expected to change
\cite{Rubi:1995}. Each deposit contains 50000 dipoles. We have also 
generated 30 DLD deposits on the same lattice by this same method, 
with $T^*=\infty$; known results for DLD (see, e.g., 
\cite{Meakin:1984}) were used to check the validity of our algorithm. On the 
other hand, comparison between these and the results for finite temperatures 
reveal the effect of dipolar interactions on DLD.

\begin{figure}
\par\columnwidth=20.5pc
\hsize\columnwidth\global\linewidth\columnwidth
\displaywidth\columnwidth
\epsfxsize=3.5truein
\centerline{\epsfig{file=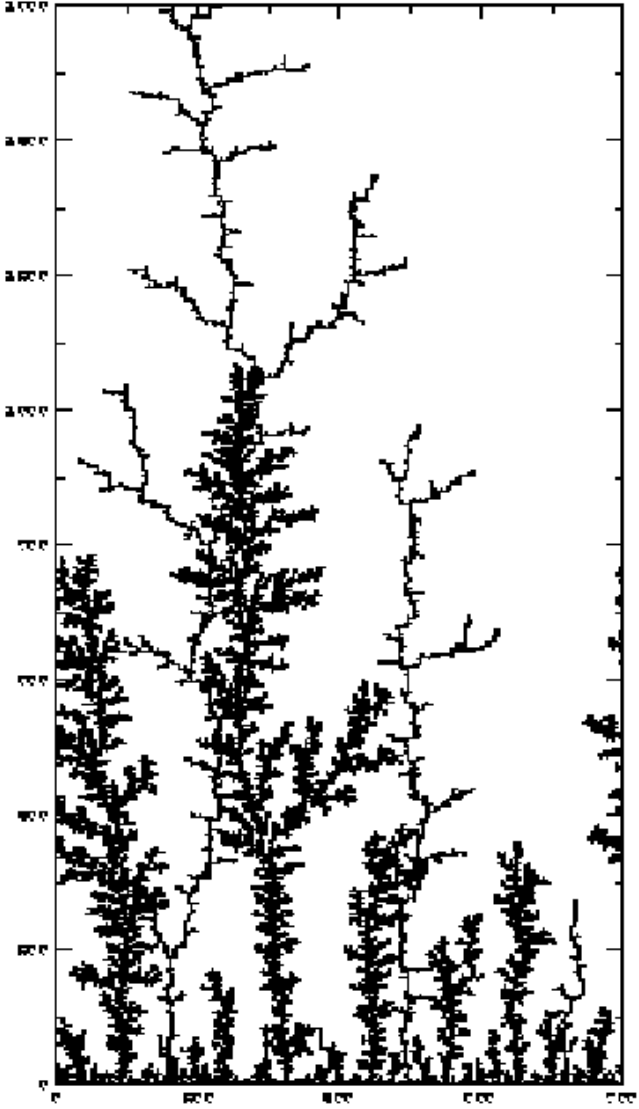,width=20pc}}
\vspace{0.5cm}
\caption{Snapshots of two deposits obtained for $T^*=10^{-1}$
(black) and $T^*=10^{-4}$ (grey).}
\label{fig1}
\end{figure}

Figure \ref{fig1} is a snaphsot of two deposits for
$T^*=10^{-1}$ (black) and $T^*=10^{-4}$ (grey, only part of the 
deposit is shown -- in fact, it grows up to a height of about 8000).
Both deposits have the same general appearance, already observed in DLD:
they consist of several trees competing to grow. As the size of the deposit
increases, fewer and fewer trees `survive' (i.e., carry on growing), as a
consequence of the so-called shielding or screening effect. From figure 
\ref{fig1} this seems more pronounced at lower temperatures, since above 
a height of 1000 (about 1/8 of the maximum height attained by this deposit) 
only one tree survives.

In order to compare quantitatively the results obtained for different
temperatures we have measured the mean density of dipoles
$\rho(h)$ at a height $h$,
\begin{equation}
\label{eq:rhodef}
\rho(h)= \langle \frac{1}{L} \sum_{i=1}^L \eta(i,h)\rangle,
\end{equation}
\begin{figure}
\par\columnwidth=20.5pc
\hsize\columnwidth\global\linewidth\columnwidth
\displaywidth\columnwidth
\epsfxsize=3.5truein
\centerline{\epsfig{file=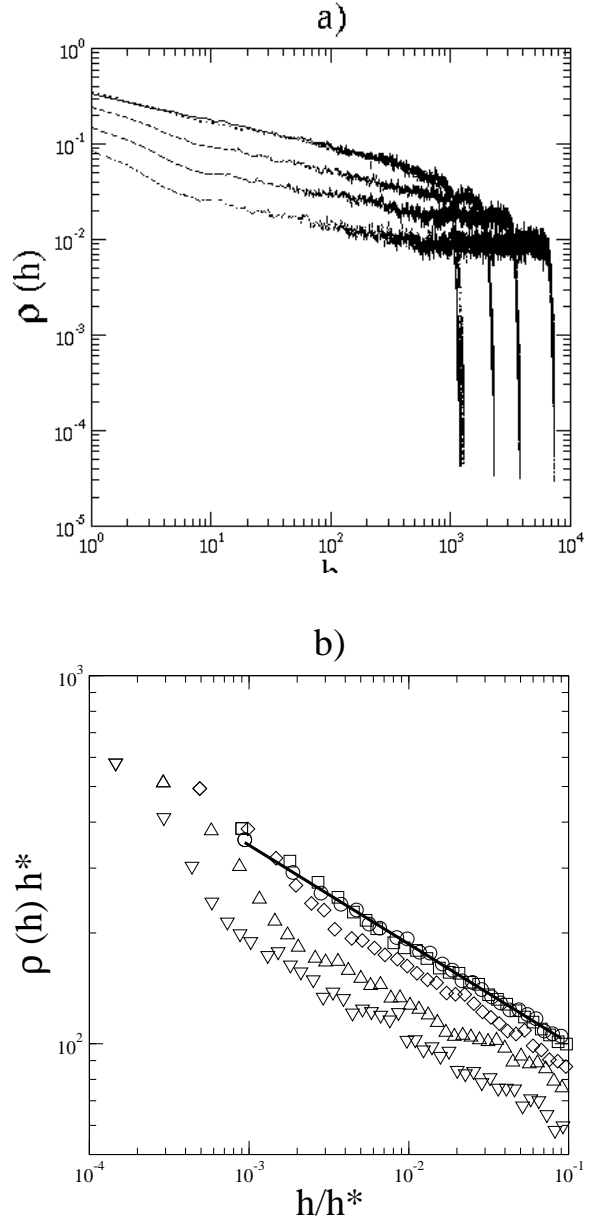,width=18pc}}
\vspace{0.5cm}
\centerline{\epsfig{file=randepdipfig2b.eps,width=18pc}}
\vspace{0.5cm}
\caption{(a) Mean density $\rho(h)$ of deposits at height $h$:
random deposition (solid line), $T^*=10^{-1}$ (dotted line), $10^{-2}$
(dashed line), $10^{-3}$ (long-dashed line) and $10^{-4}$ (dot-dashed line).
(b) $h^*\rho(h)$ {\it vs} $h/h^*$: random deposition (circles),
$T^*=10^{-1}$ (squares), $10^{-2}$ (diamonds), $10^{-3}$ (triangles up) and 
$10^{-4}$ (triangles down). The solid line is a linear fit to the DLD results
in the range of the graph; it has slope $-0.28$.}
\label{fig2}
\end{figure}
where $\eta(i,h)$ is 1 (0) if the site with coordinates $(i,h)$ is
occupied (unoccupied), and the average (denoted by angular brackets) 
is taken over all available deposits at each temperature. $\rho(h)$ 
is plotted in figure \ref{fig2}a: all the curves have similar shapes, 
with a smooth decrease at small $h$, levelling off (saturating) at 
intermediate $h$, and with a sharp drop at large $h$, when the top 
of the deposit is reached. It is immediately noticed that the 
finite-temperature curves differ from that for DLD in one 
important respect: the density at a given height and the 
maximum height $h^*$ attained by the deposits vary with 
temperature. These variations are monotonic: $\rho(h)$ 
is smaller and $h^*$ is larger at lower temperatures and thus
the increase in the strength of dipolar interactions enhances 
the shielding effect.

For DLD, $\rho(h)$ was found to be of the form \cite{Meakinbook},
\begin{equation}
\label{eq:rhohsc}
\rho(h)\propto h^{-\alpha} g(h/h^*),
\end{equation}
where $\alpha$, the so-called codimensionality, is the difference between
the dimension of space $d$ ($=2$, in the presente case) and the fractal
dimension $D$ of the deposit, $h^*$ is the maximum height, and $g(x)$
is a universal function. $g(x)\approx 1$ when $x\ll 1$ and decays faster
than any power of $x$ when $x\rightarrow 1$. In order to compare DLD and
finite-temperature results, we propose a general form for $\rho(h)$,
inspired by equation (\ref{eq:rhohsc}):
\begin{equation}
\label{eq:genrho}
\rho(h,T^*)= A(T^*) h^{-\alpha} g(h/h^*),
\end{equation}
where $A(T^*)$ is some unknown function of $T^*$ only. It is easily seen
that $A(T^*)$ can be found as a function of $h^*$ and of the number of
particles in the deposit $N$, by using the normalization condition,
\begin{equation}
\label{eq:normal}
N=L\times \sum_{h=1}^{h^*} \rho(h,T^*) \approx
L\int_{1}^{h^*} \rho(h,T^*)\,dh,
\end{equation}
whence
\begin{equation}
\label{eq:AT}
A(T^*)=\frac{N h^{*(\alpha-1)}}{L \int_{1/h^*}^1 x^{-\alpha}g(x)\,dx}.
\end{equation}
Since $g(x)$ is a universal function and $L$ and $N$ are the same for
all the deposits we are analysing, equation (\ref{eq:genrho}) becomes
\begin{equation}
\label{eq:rhoTscale}
\rho(h,T^*) h^* \propto \left(\frac{h}{h^*}\right)^{-\alpha} g(h/h^*).
\end{equation}
In figure \ref{fig2}b we plot $h^*\rho(h,T^*)$ as a function of $h/h^*$,
for $h/h^* \ll 1$ on a log-log scale. The data points for DLD follow, as
expected, a straight line. A linear regression gives $\alpha \approx 0.27$
and thus $D \approx 1.73$, in good agreement with what was obtained previously
by several other methods \cite{Meakin:1983,Meakin:1984}. If the full functional
dependence of $\rho(h,T^*)$ were captured by equation (\ref{eq:genrho}),
two possibilities would arise: (i) $\alpha$ is $T^*$-independent and all 
curves are paralell straight lines; (ii) $\alpha$ depends on $T$ and all 
curves are straight lines with different slopes. 
However, we arrive at neither of these 
scenarios, so the situation is a little more complex. If we were to 
interpret our results in terms of a function similar to equation 
(\ref{eq:genrho}), then $g(x)$ would also need to have an explicit 
temperature dependence. This dependence should be able to describe the 
trends observed in figure \ref{fig2}b for finite temperatures: a crossover 
between an approximately linear regime for very low relative heights, 
characterised by an exponent greater than $\alpha=0.27$; and a linear 
regime for intermediate heights, characterised by roughly the same 
exponent as DLD.

In \cite{Rubi:1995} it was argued that the change in the conformational
properties of DLA clusters introduced by the presence of dipolar
interactions could be interpreted as a change in their fractal dimension. 
The fractal dimension for each temperature was determined by measuring
the dependence of the radius of gyration of dipolar DLA clusters
on the number of particles in a cluster. Between $T^*=10^{-1}$ and
$T^*=10^{-4}$ a fractal dimension was obtained ranging from about 1.7
to about 1.2. We have performed linear fits to the data shown in figure
\ref{fig2}b, using only those points corresponding to heights below
the region where the crossover referred to above seems to take place.
These points follow straight lines with temperature-dependent slopes ranging
from 0.3 (for $T^*=10^{-1}$) to 0.6 (for $T^*=10^{-4}$). On the basis of
the analysis of just this part of the deposit, we obtain an (apparent)
variation of the fractal dimension of the deposits with $T^*$, from
$D=1.7$ to $D=1.4$. We conclude, as was already pointed out in
\cite{Meakinbook} and seems to be confirmed by the present work, 
that the results of \cite{Rubi:1995} can be interpreted in terms of a
crossover between a temperature-dependent fractal dimension
at short length scales, to $D\approx 1.7$ at long length scales,
with a crossover height that itself depends on temperature.
This conclusion must, however, be tested with longer simulations at
the lowest temperatures, where the statistics are a little poorer.

There are other routes to estimating the fractal dimension of the
deposits. We shall use one other to show that it is not necessary to assume
that the finite temperature deposits have a fractal dimension different
from DLD. The mean height of the upper surface, $h_m$, when the deposit 
contains $M$ particles, is defined as \cite{Meakin:1984},
\begin{equation}
\label{eq:hm}
h_m(M) =\langle \frac{1}{L} \sum_{i=1}^L h_{max}(i,M)\rangle,
\end{equation}
where $h_{max}(i,M)$ is the maximum height of the occupied sites of
column $i$ when there are $M$ particles in the deposit.
In a DLD deposit this quantity is expected to scale with $M$, as
\cite{Meakin:1984}
\begin{equation}
\label{eq:schm}
h_m \propto M^{\phi}.
\end{equation}
\begin{figure}
\par\columnwidth=20.5pc
\hsize\columnwidth\global\linewidth\columnwidth
\displaywidth\columnwidth
\epsfxsize=3.5truein
\centerline{\epsfig{file=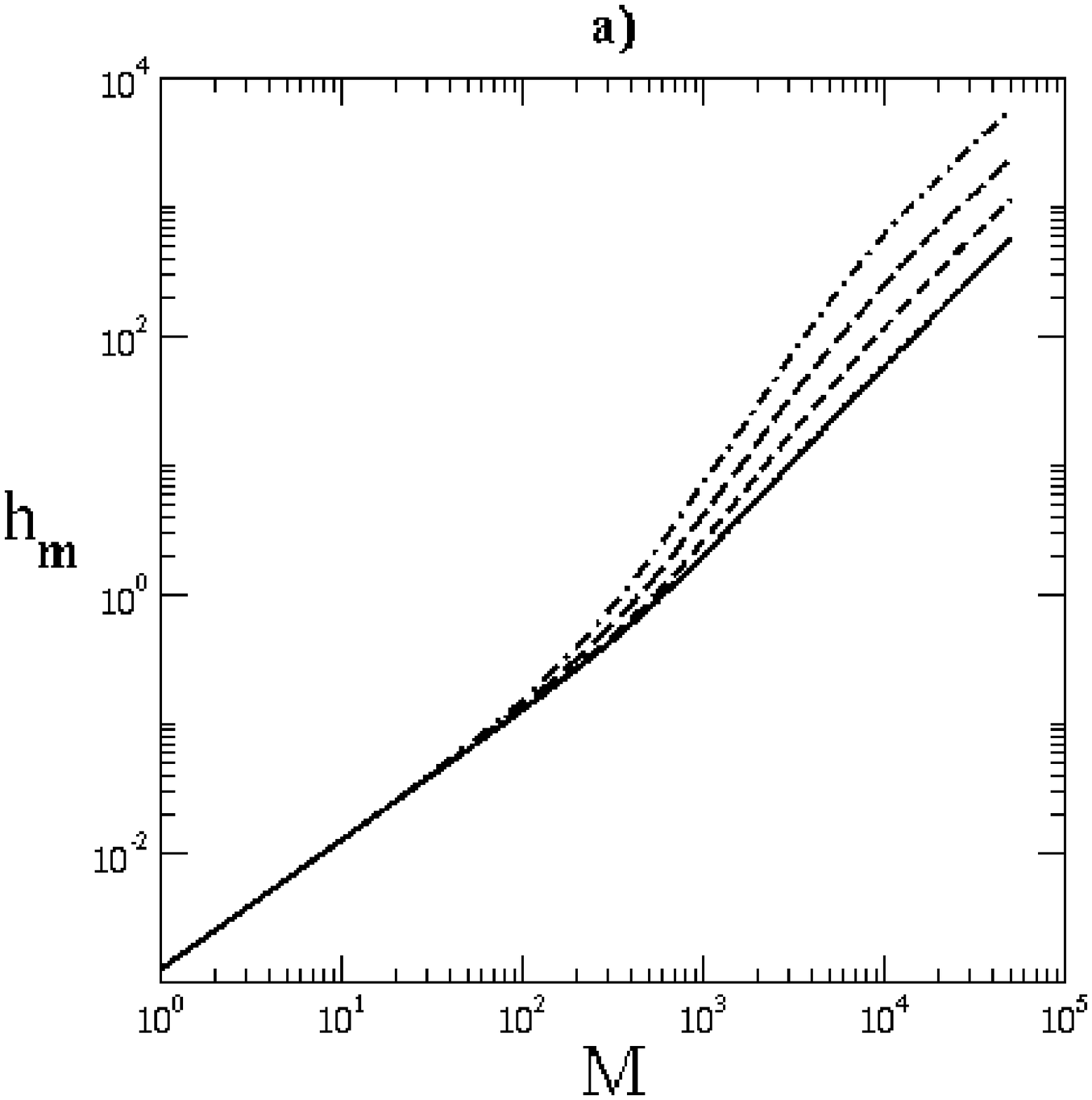,width=20pc}}
\vspace{0.5cm}
\centerline{\epsfig{file=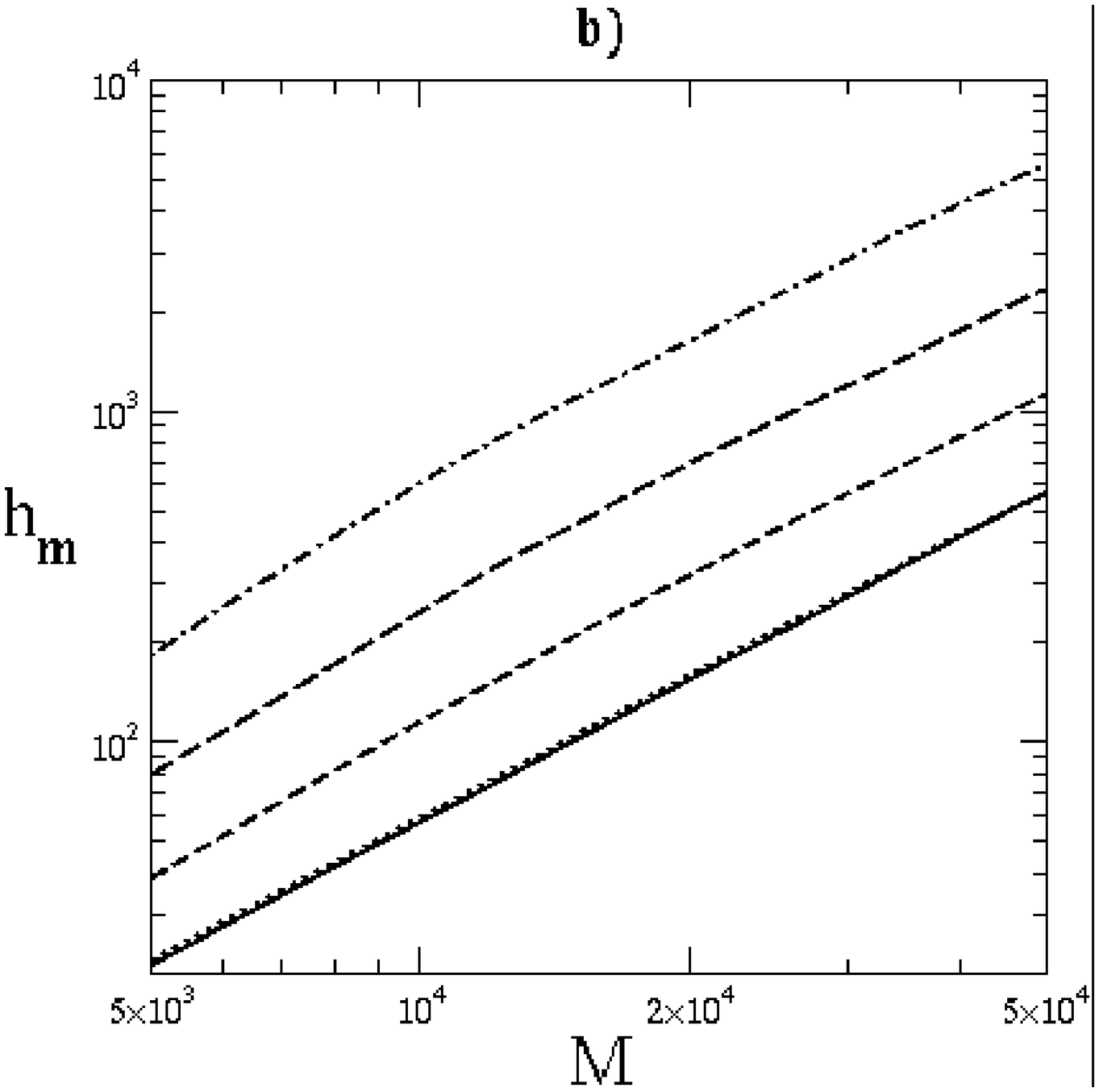,width=20pc}}
\vspace{0.5cm}
\caption{(a) Mean height of the upper surface, $h_{m}$, as a
function of the number of particles, $M$. The lines are as in figure
\protect\ref{fig2}a. (b) Blowup of the large-$M$ region.}
\label{fig3}
\end{figure}
The exponent $\phi$ is related to the codimensionality $\alpha=d-D$
by $\phi = \frac{1}{1-\alpha}$ and to the fractal dimension
by $D=d-1+\phi^{-1}$. In figure \ref{fig3} we plot our results for $h_m(M)$.
The scaling law, equation (\ref{eq:hm}), is known to be valid in the limit
$M\rightarrow \infty$ \cite{Meakin:1984}. As in \cite{Meakin:1984},
we have performed several linear regressions for large $M$, in the
range $M_{1}<M<M_{2}$, for $(M_1,M_2)=(0.5N,N)$, $(0.25N,N)$, $(0.1N,N)$
and $(0.25N,0.5N)$ (recall that $N=50000$). For every temperature and
every range considered we found that $1.33<\phi<1.44$, which corresponds 
to a fractal dimension, $1.69 <D<1.75$. Moreover, we have found no evidence 
of any regular variation of $\phi$ with $T^*$ over a given range of $M$. 
Figure \ref{fig3}a shows that, in the initial stages of growth, 
the mean height of the upper surface grows
identically at every temperature. There is then a crossover region at
intermediate stages when the less dense deposits grow slightly faster.
Finally, in the later stages all the deposits grow at the same
rate regardless of temperature, as evidenced by figure \ref{fig3}b.
However, note that, as is clear from figure \ref{fig3}a, 
if the deposits had been
allowed to grow only to intermediate stages (e.g., up to $M=10000$), 
an increasing value of $\phi$ with increasing temperature would have obtained, 
and thus an apparent variation of $D$ with $T^*$, with the same trends 
as observed through the calculation of $\rho(h)$.

\begin{figure}
\par\columnwidth=20.5pc
\hsize\columnwidth\global\linewidth\columnwidth
\displaywidth\columnwidth
\epsfxsize=3.5truein
\centerline{\epsfig{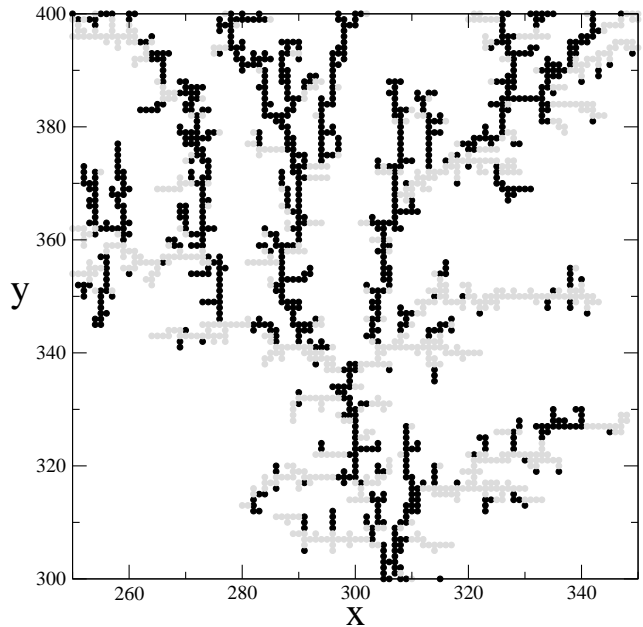}}
\vspace{0.5cm}
\caption{Detail of a deposit for $T^*=10^{-1}$. Sites whose dipoles make an
angle of absolute value smaller (larger) than $\pi/4$ with the vertical
axis are shown in black (grey).}
\label{fig4}
\end{figure}

\begin{figure}
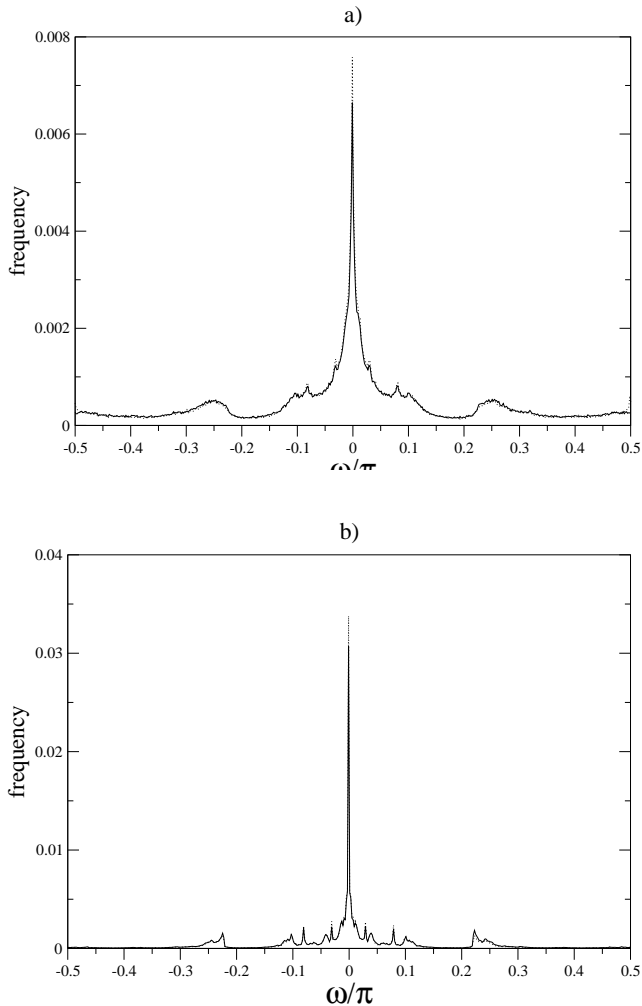

\par\columnwidth=20.5pc
\hsize\columnwidth\global\linewidth\columnwidth
\displaywidth\columnwidth
\epsfxsize=3.5truein
\vspace{0.2cm}
\centerline{\epsfig{file=randepdipfig5a.eps,width=20pc}}
\vspace{0.5cm}
\centerline{\epsfig{file=randepdipfig5b.eps,width=20pc}}
\vspace{0.5cm}
\caption{Frequency histogram of the angle $\omega$ between dipole
orientation and the local direction of growth, for (a) $T^*=10^{-1}$ and
(b) $T^*=10^{-4}$. The solid lines correspond to lateral growth and the
dotted lines to vertical growth.}
\label{fig5}
\end{figure}

We conclude this analysis by attempting to make a first connection 
between the orientation of the dipoles in the deposit and its growth. 
Figure \ref{fig4} is a snapshot of part of a deposit for $T^*=10^{-1}$.
Dipoles whose horizontal (or lateral) component is smaller (greater) in
absolute value than their vertical component are coloured black (grey).
Since we have verified that the $z$-component (i.e., out-of-plane) of
the dipoles in the deposit is zero after a short time, figure \ref{fig4}
suggests that the dipoles tend to align with the direction of growth of
the deposit at the site where they attach. 
In order to make this idea more
quantitative, we have measured the angles $\omega$ between the direction of 
the dipole moments of all incoming particles and the direction of growth at
their point of attachment to the deposit. We have done so by recording
whether a new dipole becomes attached to the substrate due to a neighbour
positioned to its left or to its right (lateral growth: the relevant angle
is that between the dipole and the horizontal axis), or above or below it 
(vertical growth: the relevant angle is now that between the dipole 
and the vertical axis). We did not take into account particles that 
attach to the deposit having both vertical and horizontal neighbours.
Once these angles were collected, for every deposit at each temperature,
we constructed a frequency histogram by dividing the interval $(-\pi,\pi)$ 
into 1000 sub-intervals. In figure \ref{fig5} we plot these results 
for lateral and vertical growth at $T^*=10^{-1}$ and $T^*=10^{-4}$.
Because all curves have period $\pi$ and are even, only the interval 
$(-\pi/2,\pi/2)$ is shown. The main feature of all the curves are the 
strong peaks at $\omega=0$, implying that most dipoles align in the 
direction of growth of the deposit. This is a consequence of the fact 
that the lowest-energy configuration of two dipoles a fixed distance 
apart is head-to-tail along the direction of the interdipole vector. 
These peaks are more pronounced at the lower temperature, and the peak 
for vertical growth is a little higher than that for lateral growth 
at both temperatures, implying that growth in the vertical direction 
is more likely to happen with dipoles aligned vertically than growth 
in the lateral direction with dipoles aligned horizontally.

The curves exhibit some other, lower, peaks. There is a broad, low peak around 
$\omega=\pi/2$ at $T^*=10^{-1}$, which corresponds to lateral growth with 
vertically-aligned dipoles (`black' horizontal branches in figure \ref{fig4}), 
or to vertical growth with horizontally-aligned dipoles (`grey' vertical
branches in figure \ref{fig4}), at high temperatures. This can be explained 
by noting that the second-lowest minimum of the interaction energy at fixed 
separation is for two antiparallel dipoles. These peaks seem to disappear at 
$T^*=10^{-4}$, suggesting that energetic effects become 
more important as the temperature is lowered.

There are several other peaks, occurring at the same angles for both lateral 
and vertical growth, whose positions  seem unaffected by changing 
the temperature. At the present stage of our work, we can only speculate as 
to their origin. We believe these peaks come from a combination of lattice 
effects and the properties of the dipolar interaction. It is actually known 
that the minimum-energy arrangement of $n(\ge 3)$ dipoles is obtained by 
placing them at the vertices of a regular $n$-sided polygon, tangent to the 
circumscribing circle. Thus four dipoles on a square lattice will minimse 
their energy by making $\pi/4$ angles with the horizontal and vertical 
axes, which might explain the peaks observed in figure \ref{fig4} at that 
angle. The remaining peaks may likewise correspond to other arrangements 
of dipoles realising other minima of the energy of sets of dipoles 
on a square lattice.

\section{Concluding remarks}

\label{sec-disc}

We have simulated the deposition of dipolar particles on a 1d substrate
using a lattice model. Our findings suggest that the fractal dimension 
of the deposits is the same as for DLD and hence unaffected by the dipolar 
interactions, but also that there is a crossover from 
temperature-dependent to temperature independent behaviour which can be 
very broad. A fuller characterisation in terms of the height-height 
correlation function, tree size distributions and density scaling with 
system size, is in progress and will be published elsewhere. These 
quantities would provide additional routes to the fractal dimension, thus 
allowing us further to verify (or falsify) our preliminary conclusions.
Growth and roughness exponents will also be calculated.

We have now started work on the off-lattice version of the present
model, so as to be free from any possible artifacts arising from the 
discretisation of space. Results so far suggest that the lattice has 
a very small effect, but we are currently somewhat limited by the very 
high computational cost of evaluating the interactions between a particle 
and all its periodic images at every step. More efficient algorithms are 
being developed to tackle this.

\section*{Acknowledgements}

Funding from the Funda\c{c}\~{a}o para  a Ci\^{e}ncia e Tecnologia 
(Portugal) is gratefully acknowledged in the form of post-doctoral fellowships
nos.\ SFRH/BPD/5654/2001 (F. de los Santos) and SFRH/BPD/1599/2000 
(M. Tasinkevych) and partial support (P. I. C. Teixeira).
We thank M. M. Telo da Gama for helpful discussions and J. J. Weis for
advice on how to treat the Ewald sums.

\appendix

\section{Ewald sums for the deposition process}

\label{sec-Ewald}

We have generalised to arbitrary dimensions a method proposed by 
Grzybowski {\it et al.} \cite{Grzybowski:2000} for evaluating Ewald 
sums. Here for simplicity we just present results for our case $d=1$
(where $d$ is the number of dimensions in which we impose periodic 
boundary conditions).
Our simulation box consists of a rectangle with a base of 
length $L$ and any height that can accommodate $N$ dipoles. The dipoles 
$\mbox{\boldmath $\mu$}_i$ are always three-dimensional vectors, 
with position vectors ${\bf R}_i, \ i=1,2,\ldots,N$. 
The simulation box is repeated in the ${\bf x}$
(horizontal) direction, giving rise to a regular lattice whose sites are
located at ${\bf n}=(n,0)L$. Let $\mbox{\boldmath $\mu$}$ and ${\bf R}$ 
be the dipole moment and position of an incoming particle. 
Then, the distance between the incoming
particle in the origin cell and another in an image cell is
${\bf r}_i \equiv {\bf R}- ({\bf R}_i+{\bf n}), \ i=1,2,\ldots,N$.
The total interaction energy between the incoming particle and the
$N$ particles in the box and their infinite replicas is
\begin{equation}
\label{eq:energy}
E=\sum_{i=1}^{N} {\sum_{\bf n}^\infty} \Bigg\{
{\mbox{\boldmath $\mu$}
\cdot \mbox{\boldmath $\mu$}_i
\over |{\bf r}_{i}+{\bf n}|^3}
-3{[\mbox{\boldmath $\mu$} \cdot ({\bf r}_{i}+{\bf n})]
[\mbox{\boldmath $\mu$}_i \cdot ({\bf r}_{i}+{\bf n})]
\over |{\bf r}_{i}+{\bf n}|^5 } \Bigg\}.
\end{equation}
According to the geometry of the system ${\bf r}_{i}=
(x_i,y_i,0)$ where $x_i$ ($y_i$) is the horizontal (vertical)
distance between the incoming particle and a particle in the deposit.
Note that the incoming particle does not interact with its own images.

Introducing the notation
\begin{eqnarray}
\label{eq:psi}
\psi({\bf r}) &=& \sum_{\bf n} {1\over |{\bf n}+ {\bf r}|^3}, \qquad \ \
{\bf r} \ne {\bf 0}, \\
\label{eq:theta}
\theta({\bf r},{\bf c}) &=& \sum_{\bf n}
{e^{-i {\bf c} \cdot ({\bf n}+ {\bf r})}\over |{\bf n}+ {\bf r}|^5},
\qquad
{\bf r} \ne {\bf 0}.
\end{eqnarray}
allows us to express the total energy as
\begin{eqnarray}
\label{eq:energy2}
E&=&\sum_{i=1}^N \mbox{\boldmath $\mu$}
\cdot \mbox{\boldmath $\mu$}_i \psi({\bf r}_i) \nonumber \\
& &\mbox{}+ 3 \sum_{i=1}^N
\Big(\mbox{\boldmath $\mu$} \cdot
\mbox{\boldmath $\nabla$}_{\bf c }\Big)
\Big(\mbox{\boldmath $\mu$}_i \cdot
\mbox{\boldmath $\nabla$}_{\bf c} \Big)
\theta({\bf r}_i,{\bf c})\vert_{{\bf c}={\bf 0}}.
\end{eqnarray}
To calculate $\psi({\bf r})$ and $\theta({\bf r},{\bf c})$ we use the
identities:
\begin{eqnarray}
\label{eq:1}
x^{-2u} &=& {1 \over \Gamma(u)}\int_0^\infty t^{u-1} e^{-xt^2} dt, \\
\label{eq:2}
\sum_{{\bf n}} e^{-t|{\bf r}+{\bf n}|^2-i{\bf c}\cdot ({\bf r}+{\bf n})}
&=&\bigg({\pi \over tL^2}\bigg)^{1/2}\sum_{\bf k} e^{i{\bf k}\cdot{\bf r}}
\nonumber \\
& &\mbox{}\times \exp \bigg( -{({\bf k}+ {\bf c})^2\over 4 t}\bigg).
\end{eqnarray}
${\bf k}=2\pi(k,0)/L$ with $k$ integer is a reciprocal-lattice vector
and ${\bf r}$ and ${\bf n}$ are as above. Equation (\ref{eq:1})
is a direct consequence of the definition of the Gamma function, while
equation (\ref{eq:2}) is a form of the Poisson summation formula for 
$d=1$, which is the dimensionality of the lattice formed by repeating 
the box.

For $\psi({\bf r})$ we set $u=3/2$, leading to
\begin{equation}
\label{eq:forpsi}
\psi({\bf r})= {1 \over \Gamma(3/2)} \sum_{\bf n} \int_0^\infty
t^{1/2} e^{-|{\bf r}+{\bf n}|^2t}dt.
\end{equation}
The sum over direct-lattice vectors converges fast for large $t$,
while that over reciprocal-lattice vectors does so for small $t$.
We therefore choose an arbitrary separation parameter $\alpha^2$
for the $t$ integration and decompose the lattice sum into two terms:
\begin{eqnarray}
\label{eq:forpsi2}
\psi({\bf r})&=& {2 \over \sqrt{\pi}} \sum_{\bf n} \int_{\alpha^2}^\infty
t^{1/2} e^{-t|{\bf r}+{\bf n}|^2}dt \nonumber \\
& &\mbox{}+ {2 \over \sqrt{\pi}}
\sum_{\bf n} \int_0^{\alpha^2} t^{1/2} e^{-t|{\bf r}+{\bf n}|^2}dt.
\end{eqnarray}
Taking into account that $-t|{\bf r}+{\bf n}|^2=-t|x+n|^2-ty^2$ and using
the Poisson summation formula, equation (\ref{eq:2}), we arrive at
\begin{eqnarray}
\label{eq:forpsi3}
\psi({\bf r})&=& {2 \over \sqrt{\pi}} \sum_{\bf n} \int_{\alpha^2}^\infty
t^{1/2} e^{-t|{\bf r}+{\bf n}|^2}dt \nonumber \\
& &\mbox{}+ {2 \sqrt{\pi} \over L}
\sum_{\bf k} e^{-i kx} \int_0^{\alpha^2}
\exp\left({-ty^2+{k^2 \over 4t}}\right)dt.
\end{eqnarray}
In the case of $\theta({\bf r},{\bf c})$ we set $u=5/2$ with the result
\begin{eqnarray}
\label{eq:fortheta}
\theta({\bf r},{\bf c}) &= &{1 \over \Gamma(5/2)} \sum_{\bf n}
e^{-i{\bf c}\cdot ({\bf r}+{\bf n})} \int_{\alpha^2}^\infty
t^{3/2} e^{-t|{\bf r}+{\bf n}|^2}dt \\
& &\mbox{}+{4 \over 3L}\sum_{\bf k} e^{ikx-iy c_y} \int_0^{\alpha^2}
t\exp\bigg({-ty^2+{k^2 \over 4t}}\bigg)dt. \nonumber
\end{eqnarray}

We now substitute $\psi({\bf r})$ and $\theta({\bf r},{\bf c}) $ into
equation (\ref{eq:energy2}) for $E$:
\begin{eqnarray}
\label{eq:energy3}
E&=&{2 \over \sqrt{\pi}} \sum_{i} {\sum_{\bf n}} \bigg\{
(\mbox{\boldmath $\mu$}
\cdot \mbox{\boldmath $\mu$}_i) I_{1/2}(\alpha,\beta) \nonumber \\
& &\mbox{}\qquad
-2\Big[\mbox{\boldmath $\mu$} \cdot({\bf r}_{i}+{\bf n}) \Big]
\Big[\mbox{\boldmath $\mu$}_i \cdot({\bf r}_{i}+{\bf n})\Big]
I_{3/2}(\alpha,\beta )\bigg\} \nonumber \\
& &\mbox{}+{2 \over L} \sum_{i}
\sum_{\bf k}\Big(\mu_y \mu_{iy}+\mu_z\mu_{iz}\Big) e^{i k x_i}
J_0(\alpha,y_i,k) \nonumber \\
& &\mbox{} -{4 \over L} \sum_{i} \sum_{\bf k}
\mu_y \mu_{iy} y_{i}^2 e^{i k x_i} J_1(\alpha,y_i,k) \nonumber \\
& &\mbox{} +{1\over L} \sum_{i} \sum_{{\bf k}\ne {\bf 0}}
\mu_x \mu_{ix} k^2 e^{i k x_i} J_{-1}(\alpha,y_i,k) \\
& &\mbox{}+i{2 \over L}\sum_{i} \sum_{{\bf k}\ne {\bf 0}}
k \Big(\mu_x \mu_{iy}+ \mu_{ix} \mu_y \Big)y_i
e^{ik x_i} J_0(\alpha,y_i,k), \nonumber
\end{eqnarray}
where $\beta=|{\bf r}_{i}+{\bf n}|$ and the integrals are given by
(see Appendix \ref{sec-integrals} for details)
\begin{eqnarray}
\label{eq:Ionehalf}
I_{1/2}\Big(\alpha,|{\bf r}_{i}+{\bf n}|\Big)&=&
{\alpha e^{-|{\bf r}_{i}+{\bf n}|^2 \alpha^2} \over |{\bf r}_{i}+{\bf n}|^2}
\nonumber \\
& &\mbox{}+ {\sqrt{\pi} \over 2 |{\bf r}_{i}+{\bf n}|^3}
{\rm erfc}\Big( \alpha |{\bf r}_{i}+{\bf n}|\Big),\\
\label{eq:Ithreehalves}
I_{3/2}\Big(\alpha,|{\bf r}_{i}+{\bf n}|\Big)&=&
{\alpha e^{-|{\bf r}_{i}+{\bf n}|^2 \alpha^2} \over |{\bf r}_{i}+{\bf n}|^2}
\bigg(\alpha^2+ {3 \over 2|{\bf r}_{i}+{\bf n}|^2}\bigg)
\nonumber \\
& &\mbox{}+ {3\sqrt{\pi} \over 4 |{\bf r}_{i}+{\bf n}|^5}
{\rm erfc}\Big( \alpha |{\bf r}_{i}+{\bf n}|\Big), \nonumber \\
\label{eq:Jnu}
J_\nu(\alpha,y_i,k)&=&\int_0^{\alpha^2} t^\nu e^{-ty_i^2-k^2/4t} dt.
\end{eqnarray}
Since $E$ is real, equation (\ref{eq:energy3}) can be further simplified 
by replacing $\exp(ikx_i)$ by $\cos(kx_i)+i\sin(kx_i)$. Note also
that it is even in $k$ and that the case $k=0$ can be treated
analytically. Therefore, no distinction is made in the code between 
simulation box images with positive and negative $k$,
and the case $k=0$ is considered separately.

\section{Evaluation of some relevant integrals}

\label{sec-integrals}

The following definitions and results will be useful:
\begin{eqnarray}
\label{eq:erf}
\int_0^\infty e^{-\nu t^2} dt&=&{1 \over 2} \sqrt{\pi \over \nu}, \\
\label{eq:erfu}
{\rm erf}(u)&\equiv& {2 \over \sqrt{\pi}} \int_0^u e^{-t^2} dt, \\
\label{eq:erfc}
{\rm erfc}(u)&\equiv&1-{\rm erf}(u)
={2 \over \sqrt{\pi}} \int_u^\infty e^{-t^2} dt.
\end{eqnarray}
Two classes of integrals need to be performed:
\begin{eqnarray}
\label{eq:Inu}
I_\nu(\alpha,\beta)&=&\int_{\alpha^2}^\infty t^\nu e^{-t\beta^2} dt, \\
\label{eq:Jnu2}
J_\nu(\alpha,a,b)&=&\int_0^{\alpha^2} t^\nu e^{-at-b/t} dt.
\end{eqnarray}
We are interested in $I_{1/2}$ and $I_{3/2}$.
They can be evaluated with the help of $I(-1/2)$, which is easy:
\begin{equation}
\label{eq:Iminonehalf}
I_{-1/2}(\alpha,\beta)={\sqrt{\pi} \over \beta} \
{\rm erfc} \Big( \alpha \beta \Big),
\end{equation}
where we have made the change of variable $t=z^2/\beta^2$.
Now the cases $\nu=1/2$ and $\nu=3/2$ can be easily found
by integrating by parts:
\begin{eqnarray}
\label{eq:Ionehalf2}
I_{1/2}(\alpha,\beta)&=&
{\alpha e^{-\beta^2 \alpha^2} \over \beta^2} + {\sqrt{\pi} \over 2
\beta^3}
{\rm erfc}\Big( \alpha \beta\Big),\\
\label{eq:Ithreehalves2}
I_{3/2}(\alpha,\beta)&=&
{\alpha e^{-\beta^2 \alpha^2} \over \beta^2}
\bigg(\alpha^2+ {3 \over 2\beta^2}\bigg)
+ {3\sqrt{\pi} \over 4 \beta^5}
{\rm erfc}\Big( \alpha \beta\Big),
\end{eqnarray}
having resorted to the changes of variables $u=t^{1/2}$ and
$u^{\prime}=-e^{-\beta^2t}$.
Turning next to $J_{\nu}$, the values of $\nu$ we are interested in
depend on the system dimensionality. For $d=1$, $\nu=-1,0,1$ and no 
analytical results are available, in contrast with the $d=2$ case. One of 
the three integrals, however, can be expressed in terms of the other two:
\begin{equation}
\label{eq:JJJ}
{k^2 \over 4}J_{-1}+J_0-y^2J_1=\alpha^2
\exp\bigg(-y^2\alpha^2- {k^2 \over 4t}\bigg).
\end{equation}

\end{multicols}

\end{document}